\documentclass[aps,twocolumn,superscriptaddress]{revtex4}
\usepackage{graphics}
\usepackage{graphicx}
\usepackage{times}
\usepackage{multirow}
\usepackage{hyperref}
\hypersetup{colorlinks=false,pdfborder=0}
\usepackage{color}
\usepackage{cancel}
\hypersetup{
   bookmarksnumbered,
   colorlinks={true},
   pdfstartview={FitH},
   citecolor={black},
   linkcolor={black},
   urlcolor={black},
}

\newcommand{\wpmark}{*}
\newcommand{\memark}{$^+$}

\begin{document}

\title{Inheritance patterns in citation networks reveal scientific memes}

\author{Tobias Kuhn}
\thanks{Electronic address: \href{mailto:tokuhn@ethz.ch}{\textcolor{blue}{tokuhn@ethz.ch}}}
\affiliation{Chair of Sociology, in particular of Modeling and Simulation, ETH Zurich, 8092 Zurich, Switzerland}

\author{Matja{\v z} Perc}
\affiliation{Faculty of Natural Sciences and Mathematics, University of Maribor, Koro{\v s}ka cesta 160, SI-2000 Maribor, Slovenia}
\affiliation{CAMTP -- Center for Applied Mathematics and Theoretical Physics, University of Maribor, Krekova 2, SI-2000 Maribor, Slovenia}

\author{Dirk Helbing}
\affiliation{Chair of Sociology, in particular of Modeling and Simulation, ETH Zurich, 8092 Zurich, Switzerland}
\affiliation{Risk Center, ETH Zurich, 8092 Zurich, Switzerland}

\begin{abstract}\noindent
Memes are the cultural equivalent of genes that spread across human culture by means of imitation. What makes a meme and what distinguishes it from other forms of information, however, is still poorly understood. Our analysis of memes in the scientific literature reveals that they are governed by a surprisingly simple relationship between frequency of occurrence and the degree to which they propagate along the citation graph. We propose a simple formalization of this pattern and we validate it with data from close to 50 million publication records from the Web of Science, PubMed Central, and the American Physical Society. Evaluations relying on human annotators, citation network randomizations, and comparisons with several alternative approaches confirm that our formula is accurate and effective, without a dependence on linguistic or ontological knowledge and without the application of arbitrary thresholds or filters.
\end{abstract}

\maketitle

\noindent The evaluation of scientific output and the study of patterns of scientific collaboration have received increasing attention by researchers. From citation distributions \cite{redner_epjb98, radicchi_pnas08}, coauthorship networks \cite{newman_pnas04} and the formation of research teams \cite{guimera_s05, milojevic_pnas14}, to the ranking of researchers \cite{hirsch_pnas05, radicchi_pre09, petersen_pre10} and the quantification and prediction of scientific success \cite{wang2013science,penner_sr13} --- how we do science has become a science in its own right. While the famous works of Derek de Solla Price \cite{price_sci65} and Robert Merton \cite{merton_sci68} from the mid 1960s marked the beginning of a popular and long-lasting research field, the rapid progress made in recent years is largely due to the increasing availability of vast amounts of digitized data.
Massive publication and citation databases, also referred to as ``metaknowledge'' \cite{evans_s11}, along with leaps of progress in the theory and modeling of complex systems, fuel large-scale explorations of human culture that were unimaginable even a decade ago \cite{michel_s11}. The ``science of science'' is scaling up massively as well, with studies on global citation and collaboration networks \cite{pan_sr12}, the ``scientific food web'' \cite{mazloumian_scirep13}, and phylomemetic patterns in the evolution of science \cite{chavalarias_pone13}, culminating in the visually compelling atlases of science \cite{borner_10} and knowledge \cite{borner_14}.

Science is a key pillar of modern human culture, and the general concept of memes has proved to be very insightful for the study of culture. The term ``meme'' was coined by Richard Dawkins in his book \emph{The Selfish Gene} \cite{dawkins_89}, where he argues that cultural entities such as words, melodies, recipes, and ideas evolve similarly as genes, involving replication and mutation but using human culture instead of the gene pool as their medium of propagation. Recent research on memes has enhanced our understanding of the dynamics of the news cycle \cite{leskovec_acm09}, the tracking of information epidemics in the blogspace \cite{adar_ieee05}, and the political polarization on Twitter \cite{conover_aaai11}. It has been shown that the evolution of memes can be exploited effectively for inferring networks of diffusion and influence \cite{gomez_kdd10}, and that information contained in memes is evolving as it is being processed collectively in online social media \cite{simmons_aaai11}.
The question of how memes compete with each other for the limited and fluctuating resource of user attention has also amassed the attention of scientists, demonstrating that social network structure is crucial to understand the diversity of memes \cite{weng2012sr}, which suggests that social contagion mechanisms \cite{christakis2013statmed} play an important role. It has also been shown that the competition among memes can bring the network at the brink of criticality \cite{stanley_71}, where even minute disturbances can lead to avalanches of events that make a certain meme go viral \cite{gleeson_prl14}.

While the study of memes in mass media and popular culture has been based primarily on their aggregated bursty occurrence patterns, the citation network of scientific literature allows for more sophisticated and fine-grained analyses. \emph{Quantum}, \emph{fission}, \emph{graphene}, \emph{self-organized criticality}, and \emph{traffic flow} are examples of well-known memes from the field of physics, but what exactly makes such memes different from other words and phrases found in the scientific literature? As an answer to this question, we propose the following definition that is a modified version of Dawkins' definition of the word ``gene'' \cite{dawkins_89}: \emph{A scientific meme is a short unit of text in a publication that is replicated in citing publications and thereby distributed around in many copies; the more likely a certain sequence of words is to be broken apart, altered, or simply not present in citing publications, the less it qualifies to be called a meme.} Publications that reproduce words or phrases from cited publications are thus the analogue to offspring organisms that inherit genes from their parents.
In contrast to existing work on scientific memes, our approach is therefore grounded in the ``inheritance mechanisms'' of memes and not just their accumulated frequencies.
The above definition covers memes made up of exact words and phrases, but the same methods apply just as well to more abstract forms of memes.

\begin{figure*}[tb]
\centering{\includegraphics[width = 17cm]{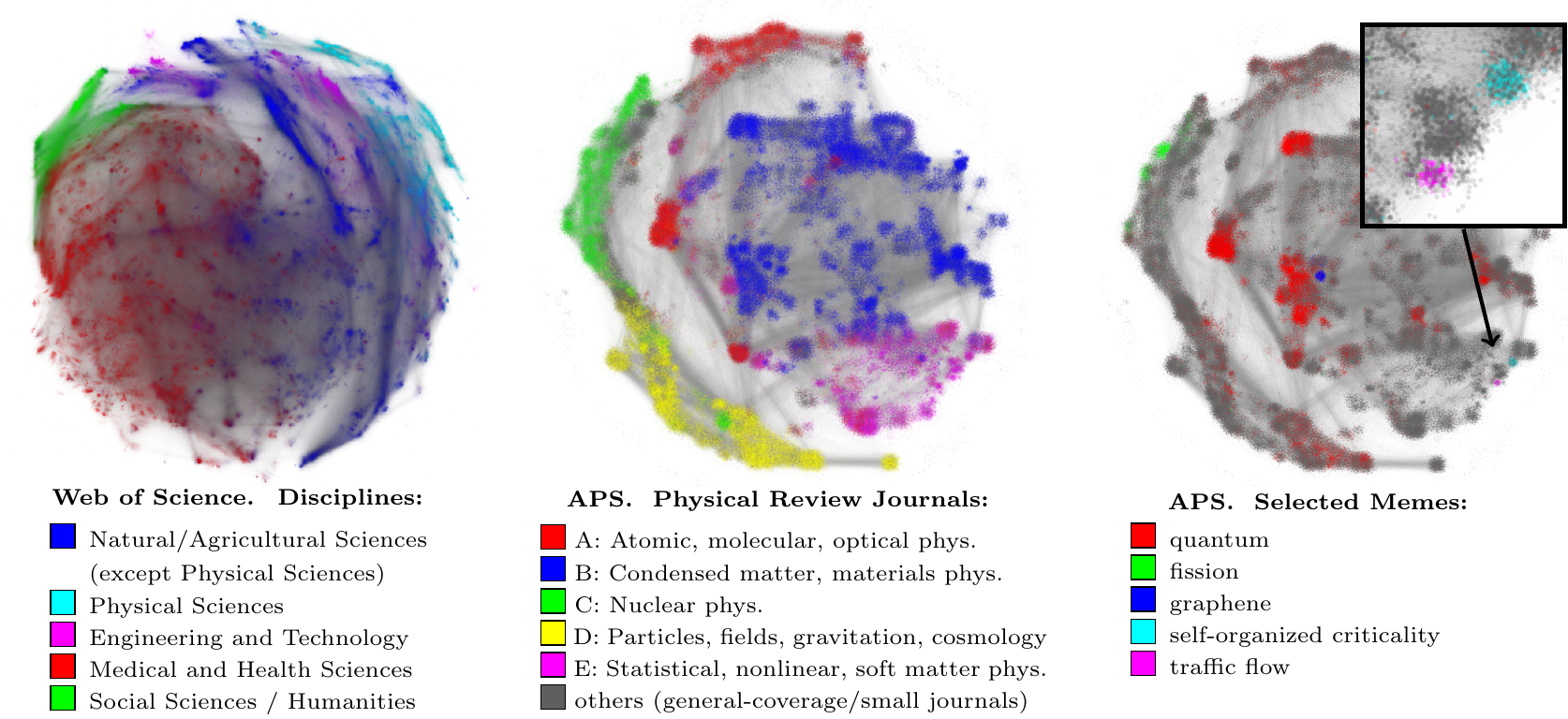}}
\caption{
\label{citgraph}
Citation networks of the Web of Science and the American Physical Society (APS) datasets reveal community structures that nicely align with scientific disciplines, journals covering particular subfields, and occurrences of memes. The generation of the visualizations was based on Gephi \cite{bastian2009icwsm} and the OpenOrd plugin \cite{martin2011spie}, which implements a force-directed layout algorithm that is able to handle very large graphs.
}
\end{figure*}

For our analyses, we rely on 47.1 million publication records from three sources. Due to its representative long-term coverage of a specific field of research, we focus mainly on the titles and abstracts from the dataset of the American Physical Society, consisting of almost half a million publications from the Physical Review journals published between July 1893 and December 2009. We also present results for the over 46 million publications indexed by the comprehensive Web of Science database, and for the over 0.6 million publications from the open access subset of PubMed Central that covers research mostly from the biomedical domain and mostly from recent years.

Fig.~\ref{citgraph} shows visualizations of these citation graphs. The leftmost network depicts the entire giant component of the citation graph of the Web of Science, consisting of more than 33 million publications. Different scientific disciplines form relatively compact communities: The physical sciences (cyan) are close to engineering and technology (magenta) in the top right corner of the network, but rather far from the social sciences and humanities (green) as well as the medical and health sciences (red), which take up the majority of the left hand side of the network, with the natural and agricultural sciences in between (blue).
Zooming in on the physical sciences and switching to the dataset from the American Physical Society, we get the picture shown in the middle. The colors now encode the five most important special-focus journals of Physical Review, each covering a particular subfield of physics (general-coverage and smaller journals are shown in gray). We see a complex structure with many small and large clusters.
Importantly, even though the employed layout algorithm \cite{martin2011spie} did not take the scientific disciplines and the journal information explicitly into account, the different communities can be clearly inferred in the citation graphs.
Following our general meme-centric perspective, the rightmost network highlights the above-mentioned memes from physics, which mostly appear in publications that form compact communities in the citation graph. The meme \emph{quantum} is widely but by no means uniformly distributed, pervading several large clusters. Publications containing the meme \emph{fission} form a few connected clusters limited to an area that makes up the journal Physical Review C covering nuclear physics. Similarly, the memes \emph{graphene}, \emph{self-organized criticality}, and \emph{traffic flow} (see enlarged area) are each concentrated in their own medium-sized or small communities.

\section*{Results}

All words and phrases that occur frequently in the literature can be considered important memes, but many frequent words like ``method'' are not particularly interesting for any given scientific field. To quantify the degree to which a meme is interesting, we define the propagation score $P_m$, which determines the alignment of the occurrences of a given meme with the citation graph.
$P_m$ is high for memes that frequently appear in publications that cite meme-carrying publications (``sticking'') but rarely appear in publications that do not cite a publication that already contains the meme (``sparking''). Formally, we define the propagation score for a given meme $m$ as its \emph{sticking factor} divided by its \emph{sparking factor}. The sticking factor quantifies the degree to which a meme replicates in a publication that cites a meme-carrying publication.
Concretely, it is defined as $d_{m{\rightarrow}m} / d_{{\rightarrow}m}$, where $d_{m{\rightarrow}m}$ is the number of publications that carry the meme and cite at least one publication carrying the meme, while $d_{{\rightarrow}m}$ is the number of all publications (meme-carrying or not) that cite at least one publication that carries the meme.
Similarly, the sparking factor quantifies how often a meme appears in a publication without being present in any of the cited publications. It is thus defined as $d_{m{\rightarrow}\cancel{m}} / d_{{\rightarrow}\cancel{m}}$, where $d_{m{\rightarrow}\cancel{m}}$ is the number of meme-carrying publications that do \emph{not} cite publications that carry the meme, and $d_{{\rightarrow}\cancel{m}}$ is the number of \emph{all} publications (meme-carrying or not) that do \emph{not} cite meme-carrying publications. For the propagation score $P_m$, we thus obtain
\begin{equation}
P_m = \left. \frac{d_{m{\rightarrow}m}}{d_{{\rightarrow}m}} \middle/ \frac{d_{m{\rightarrow}\cancel{m}}}{d_{{\rightarrow}\cancel{m}}} \right. .
\label{pscore}
\end{equation}
Based on the propagation score $P_m$ and the frequency of occurrence $f_m$ (which is simply the ratio of publications carrying the meme) of a particular meme $m$, we define the \emph{meme score} $M_m$ as
\begin{equation}
M_m = f_m P_m.
\label{mscore}
\end{equation}

The propagation score, as defined in Eq.~\ref{pscore}, can be improved by adding a small amount of controlled noise $\delta$, thus obtaining
\begin{equation}
P_m = \left. \frac{d_{m{\rightarrow}m}}{d_{{\rightarrow}m}+\delta} \middle/ \frac{d_{m{\rightarrow}\cancel{m}}+\delta}{d_{{\rightarrow}\cancel{m}}+\delta} \right. .
\end{equation}
This corrects for the fact that any of the four basic terms can be zero, and it also prevents that phrases with a very low frequency get a high score by chance.
The controlled noise corresponds to $\delta$ fictitious publications that carry all memes and cite none, plus another $\delta$ publications that carry no memes and cite all. This decreases the sticking factors and increases the sparking factors of all memes, thereby reducing all meme scores --- very slightly so for frequent memes but heavily for rare ones. Our tests show that a small value of $\delta$ (e.g. $\delta=3$ as used throughout this work unless stated otherwise) is sufficient.
Another matter that deserves attention is the potential ``free-riding'' of shorter memes on longer ones.
For example, the multi-token meme ``the littlest Higgs model,'' contains the specific token ``littlest'' that rarely occurs otherwise. The meme ``littlest'' therefore gets about the same propagation score as the long meme, yet the larger meme is clearly more interesting. This can be addressed by discounting for free-riding by redefining the term $d_{m{\rightarrow}m}$ in Eq.~\ref{pscore} to exclude publications where the given meme appears in the publication and its cited publications only within the same larger meme. If ``littlest,'' for example, is always followed by ``Higgs'' in a given publication and all its cited publications, then this publication shall not contribute to the $d_{m{\rightarrow}m}$ term for $m = \mbox{``littlest''}$.

The meme score considers whether a meme is important ($f_m$) and whether it is interesting ($P_m$), and it has additionally a number of desirable properties: (i) it can be calculated exactly without the introduction of arbitrary thresholds, such as a minimum number of occurrences, without limiting the length of $n$-grams to consider, and without filtering out words containing special characters; (ii) it does not depend on external resources, such as dictionaries or other linguistic data; (iii) it does not depend on filters, like stop-word lists, to remove the most common words and phrases; and (iv) it is very simple with only one parameter ($\delta$).

\begin{figure*}
\centering{\includegraphics[width = 15.5cm]{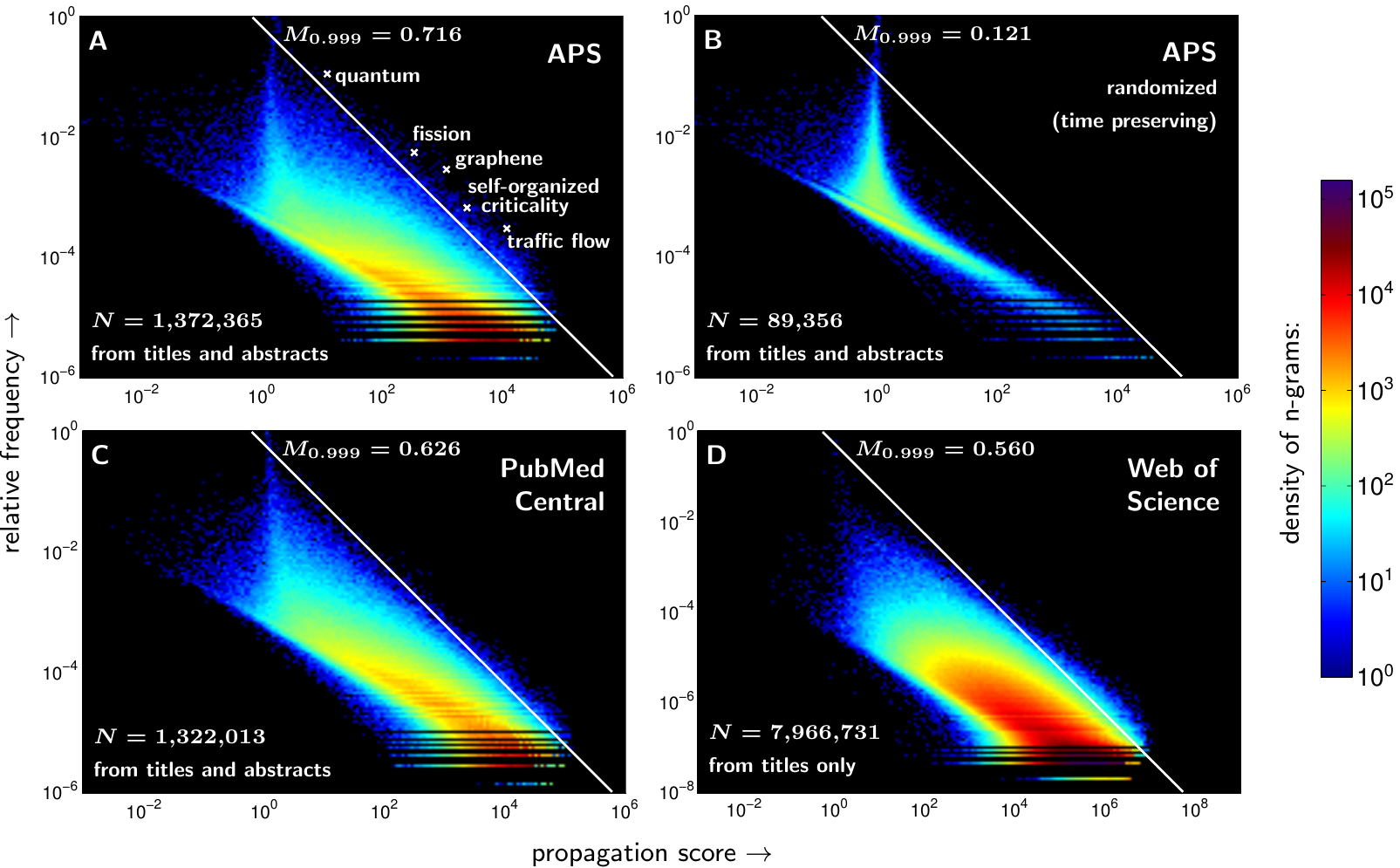}}
\caption{
\label{score}
Universality in the distribution patterns of scientific memes across datasets. Heat maps encode the density of all $n$-grams with $M \neq 0$ ($N$ being the number of such $n$-grams) with respect to their propagation score and frequency.
Maps A, C and D each show a broad band with a downward slope for the datasets from the American Physical Society (APS), the open access subset of PubMed Central, and the Web of Science, respectively. The 99.9\%-quantile with respect to the meme score distribution ($M_{0.999}$) is depicted as a white line. Memes are located mostly around the very edge of the top-right side of the band (in the vicinity of the 99.9\%-quantile line). Heat map B shows the results obtained with a time-preserving randomization of the APS citation graph (see Methods for details).
}
\end{figure*}

Calculating the meme score for all $n$-grams in the three datasets considered gives us the results presented in Fig.~\ref{score}. Their relatively frequency and their propagation score are plotted against each other in the form of heat maps with logarithmic scales. There is no upper limit to the length of $n$-grams, and the presented maps cover without exception all $n$-grams with a non-zero meme score ($N$ being the number of such $n$-grams). Meme scores are increasing towards the top-right and decreasing towards the bottom-left corner. Maps A, C and D feature a broad band with a downward slope, indicating that, in general, more frequent memes tend to propagate less via the citation graph. In the lower half of each map, we see a wedge of very high densities that follows the larger band on the bottom-left edge, but getting narrower towards the middle where it ends. Though this wedge has a somewhat rounder and broader shape for the Web of Science database, overall these patterns look remarkably similar across all datasets despite their differences with respect to topic, coverage, and size. This is an indication of universality in the distribution patterns of scientific memes. The 99.9\%-quantile line ($M_{0.999}$) is also surprisingly stable, considering that the underlying values range over five orders of magnitude or more. Localizing the previously mentioned physics memes in the APS dataset (map A), we see that they are located on the very edge of the top-right side of the band, where the density of $n$-grams is very low.
(Very frequent words like ``of'' or ``the'' are found in the faint spike at the top of the plot where $P \approx 1$ and the frequency is close to 100\%.)

The heat map B in Fig.~\ref{score} illustrates a typical case of what happens when the APS citation graph is randomized but the time ordering of publications is preserved. The number of terms with a non-zero meme score decreases dramatically (from $\approx\,$1.4 million in map A to just 89,356 in map B), the universal distribution pattern of scientific memes vanishes, and the top-right part, where the top-ranked memes should be located, disappears completely. Naturally, if the APS citation graph is randomized without preserving the time ordering, the overlap with the original results presented in map A is even smaller (see Supplementary Material). Statistical analysis reveals that median values of the meme score obtained with the randomized networks differ by more than one order of magnitude from those obtained with the original citation graph, with very little variation between different randomization runs. These results show that topology and time structure alone fail to account for the reported universality in the distribution patterns, and that thus the top memes get their high meme scores based on intricate processes and conventions that underlie the dynamics of scientific progress and the way credit is given to previous work.

Table~\ref{top50} shows the $50$ top-ranked memes from the APS dataset, also indicating their agreement with human annotation and whether they can be found under a subcategory of physics in Wikipedia. Most of these memes are noun phrases denoting real and reasonable physics concepts, which is remarkable given that the computation of the simple meme score formula uses no linguistic or ontological knowledge whatsoever.
The dominance of noun phrases is consistent with the finding that (scientific) concepts are typically captured by noun phrases when represented as keywords in terminologies \cite{bourigault1996euralex,hulth2003emnlp}.
The extracted memes consist of one, two or three words, which indicates that the meme score does not favor short or long phrases, again without applying explicit measures to balance $n$-gram lengths.
A further observation is that chemical formulas such as MgB$_2$ and CuGeO$_3$ are relatively frequent, which we investigate in more detail below.

\begin{table*}[ht]
\small
\begin{tabular*}{\hsize}{@{\extracolsep{\fill}}r@{}l@{~}r@{}l@{~}r@{}l@{~}r@{}l}
1. & loop quantum cosmology\memark\wpmark & 14. & strange nonchaotic & 27. & Na$_{x}$CoO$_{2}$\memark & 38. & inspiral\wpmark \\
2. & unparticle\memark\wpmark & 15. & in NbSe$_{3}$ & 28. & the unparticle\memark & 39. & spin Hall effect\memark\wpmark \\
3. & sonoluminescence\memark\wpmark & 16. & spin Hall\memark & 29. & black & 40. & PAMELA \\
4. & MgB$_{2}$\memark & 17. & elliptic flow\memark\wpmark & 30. & electromagnetically induced & 41. & BaFe$_{2}$As$_{2}$\memark \\
5. & stochastic resonance\memark\wpmark & 18. & quantum Hall\memark\wpmark &  & transparency\memark\wpmark & 42. & quantum dots\memark\wpmark \\
6. & carbon nanotubes\memark\wpmark & 19. & CeCoIn$_{5}$\memark & 31. & light-induced drift\memark & 43. & Bose-Einstein condensates\memark \\
7. & NbSe$_{3}$\memark & 20. & inflation\memark & 32. & proton-proton bremsstrahlung\memark & 44. & X(3872)\wpmark \\
8. & black hole\memark\wpmark & 21. & exchange bias\memark\wpmark & 33. & antisymmetrized molecular & 45. & relaxor\memark \\
9. & nanotubes\memark & 22. & Sr$_{2}$RuO$_{4}$\memark & & dynamics\memark & 46. & blue phases\memark \\
10. & lattice Boltzmann\memark\wpmark & 23. & traffic flow\memark\wpmark & 34. & radiative muon capture\memark & 47. & black holes\memark\wpmark \\
11. & dark energy\memark\wpmark & 24. & TiOCl & 35. & Bose-Einstein\memark & 48. & PrOs$_{4}$Sb$_{12}$\memark \\
12. & Rashba & 25. & key distribution\memark & 36. & C$_{60}$\memark & 49. & the Schwinger multichannel method\memark \\
13. & CuGeO$_{3}$\memark & 26. & graphene\memark\wpmark & 37. & entanglement\memark & 50. & Higgsless\memark \\
\end{tabular*}
\caption{
\label{top50}
Top 50 memes with respect to their meme score from the APS dataset. The symbol {\memark} indicates memes where the human annotators agreed that this is an interesting and important physics concept, while the symbol {\wpmark} indicates memes that are also found on the list of memes extracted from Wikipedia (see Methods for details).}
\end{table*}

\begin{table*}[tb]
\begin{center}\small
\begin{tabular}{l@{\hspace{10mm}}rcc|cc|cc}
\textbf{method} & main class & ~ annotator & (annotator ~ & \multicolumn{2}{c|}{classified as main class:} & \multicolumn{2}{c}{$p$-value for difference to:} \\
 & & & agreement) & \emph{~ individually ~} & \emph{~ in agreement ~} & \emph{~ ~ ~ random ~ ~ ~} & \emph{weighted random} \\
\hline
 & \multirow{2}{*}{physics concept} & A1 & \multirow{2}{*}{(90.0\%)} & 85.3\% & \multirow{2}{*}{81.3\%} & \multirow{2}{*}{${<}10^{-15}$*} & \multirow{2}{*}{${<}10^{-15}$*} \\
\multirow{2}{*}{\textbf{meme score}} & & A2 & & 87.3\% & & & \\
\cline{3-8}
 & \multirow{2}{*}{noun phrase} & A1 & \multirow{2}{*}{(93.3\%)} & 86.0\% & \multirow{2}{*}{82.7\%} & \multirow{2}{*}{${<}10^{-15}$*} & \multirow{2}{*}{${<}10^{-15}$*} \\
 & & A2 & & 86.0\% & & & \\
\hline
 & \multirow{2}{*}{physics concept} & A1 & \multirow{2}{*}{(85.3\%)} & 32.7\% & \multirow{2}{*}{25.3\%} & \multirow{2}{*}{--} & \multirow{2}{*}{0.123} \\
\multirow{2}{*}{\textbf{random}} & & A2 & & 32.7\% & & & \\
\cline{3-8}
 & \multirow{2}{*}{noun phrase} & A1 & \multirow{2}{*}{(86.0\%)} & 39.3\% & \multirow{2}{*}{33.3\%} & \multirow{2}{*}{--} & \multirow{2}{*}{0.163} \\
 & & A2 & & 36.0\% & & & \\
\hline
 & \multirow{2}{*}{physics concept} & A1 & \multirow{2}{*}{(90.7\%)} & 20.7\% & \multirow{2}{*}{19.3\%} & \multirow{2}{*}{0.123} & \multirow{2}{*}{--} \\
\multirow{2}{*}{\textbf{weighted random}} & & A2 & & 27.3\% & & & \\
\cline{3-8}
 & \multirow{2}{*}{noun phrase} & A1 & \multirow{2}{*}{(86.0\%)} & 28.7\% & \multirow{2}{*}{25.3\%} & \multirow{2}{*}{0.163} & \multirow{2}{*}{--} \\
 & & A2 & & 28.7\% & & & \\
\end{tabular}
\caption{The two annotators (A1 and A2) classified more than 80\% of the memes with the highest meme scores as relevant physics concepts and noun phrases. The differences involving the meme score are highly significant (*). See Methods and Supplementary Material for details.}
\label{human}
\end{center}
\end{table*}

In Table~\ref{human}, we present results of a manual annotation of terms identified by meme score as compared to randomly selected terms (see also Supplementary Material).
Each of the annotators considered around 86\% of the meme score terms to be important physics concepts, agreeing on this in 81\% of the cases. With respect to their linguistic categories, each annotator considered 86\% of the meme score terms to be noun phrases, and the two annotators agreed on that for 83\% of the terms. The respective values are much lower for the randomly extracted terms. Only 25\% (non-weighted) and 19\% (weighted) of terms were, in agreement, found to be important physics concepts, and only 33\% (non-weighted) and 25\% (weighted) to be noun phrases. The reported differences between the meme score and the two random selection methods are highly significant ($p < 10^{-15}$ using Fisher's exact test on the number of agreed classifications). These results confirm that the meme score strongly favors noun phrases and important concepts.

\begin{figure}
\centering{\includegraphics[width = 8.5cm]{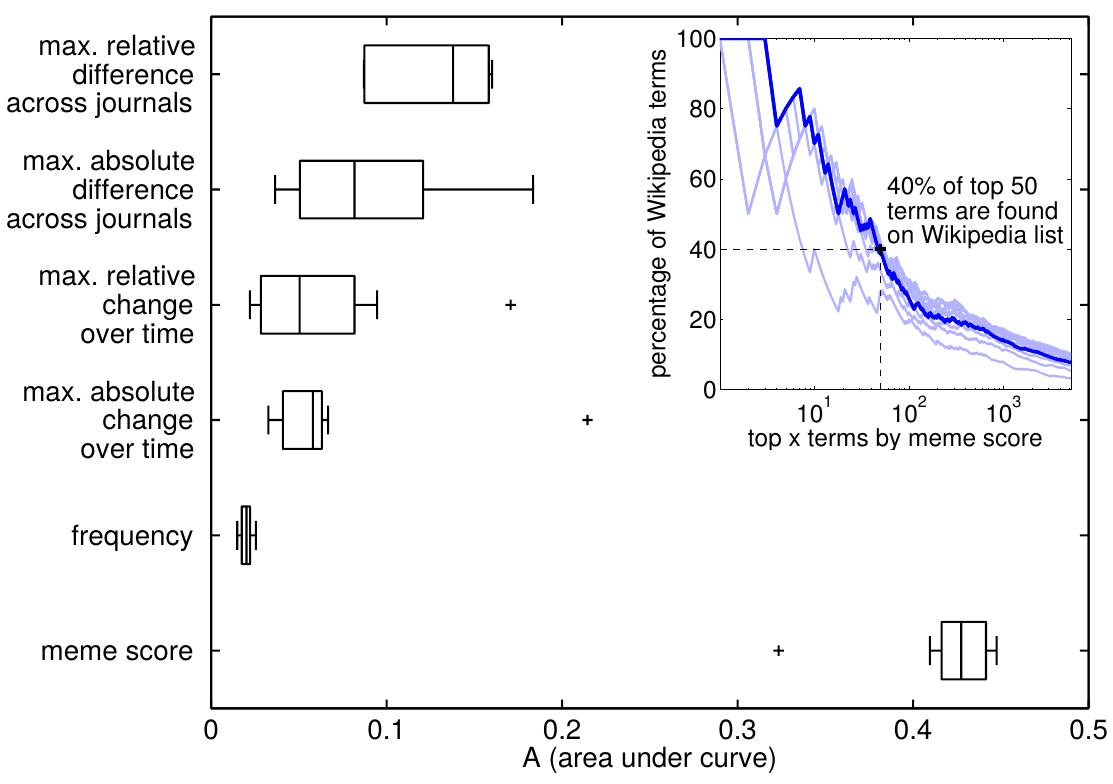}}
\caption{
\label{wiki}
The meme score outperforms alternative metrics.
The box plot shows the agreement with the ground-truth list of physics terms extracted from Wikipedia as achieved by the different metrics. Agreement is measured as the area under the curve $A$, as shown for the meme score in the embedded graph on the top right. The curves are defined as the percentage from the $x$ top-ranked terms that also appear on the ground-truth list for the different parameter settings ($1 \leq \delta \leq 10$ for the meme score; the thick line stands for $\delta=4$, which has the largest area $A$; see Methods for details).
}
\end{figure}

Next we compare the meme score to a number of possible alternative metrics, as defined in the Methods section, and align the identified words and phrases with a ground-truth list of terms extracted from physics-related Wikipedia titles.
Fig.~\ref{wiki} summarizes the results, showing on the top right that $\approx\,$70\% of the top $10$ memes identified by meme score correspond to terms extracted from Wikipedia, $\approx\,$55\% of the top $20$, $\approx\,$40\% of the top $50$, and $\approx\,$26\% of the top $100$. The largest area under the curve $A$ is obtained for a controlled noise level $\delta=4$, which is highlighted by the thick blue line.
The box plot compares the outcomes of different metrics with respect to $A$, as described in the Methods section.
The isolated outlier at 32.3\% on the lower end of the distribution for the meme score originates from the parameter setting $\delta = 1$. All other parameter settings $2 \leq \delta \leq 10$ lead to results in a narrow band between 40.9\% and 44.8\% (meaning that the performance of the metric is not sensitive to movements within this range of the parameter space). In contrast, all alternative metrics score considerably worse, consistently below 22\% (including outliers).

\begin{figure*}
\centering{\includegraphics[width = 0cm]{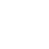}}
\centering{\includegraphics[width = 15.5cm]{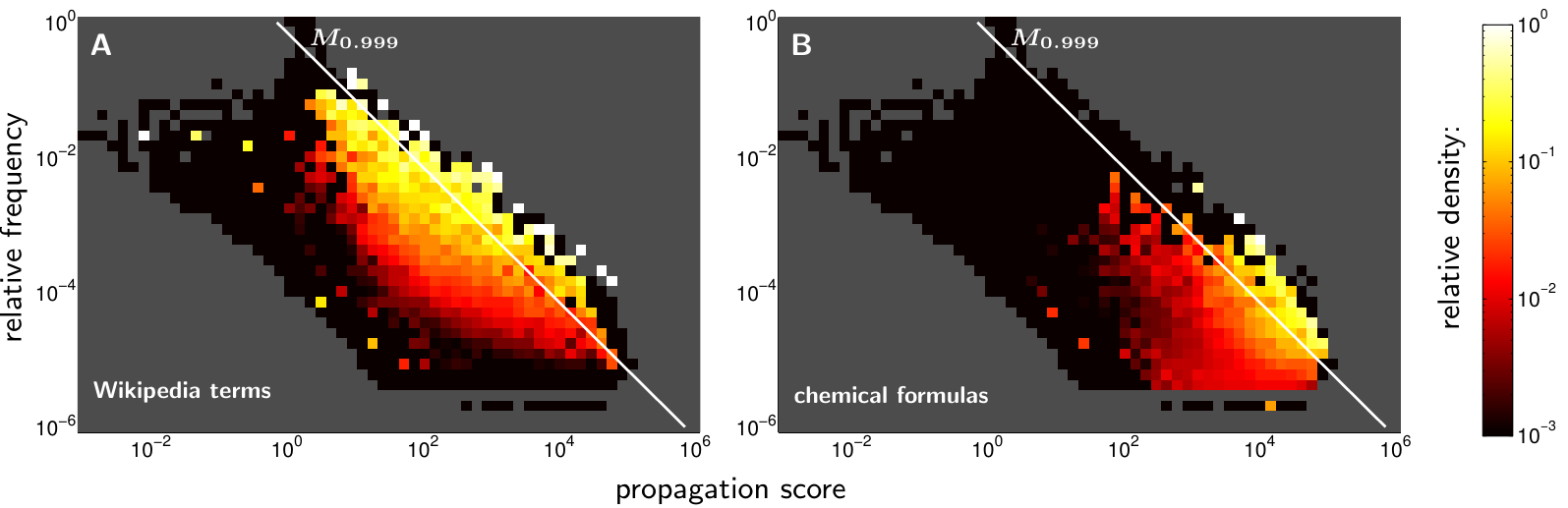}}
\caption{
\label{dens}
Plot A shows that phrases with a high meme score (i.e. around the 99.9\%-quantile $M_{0.999}$) tend to be found as titles of Wikipedia articles on physics, while other phrases tend not to. Plot B shows that phrases of relatively low frequency but high meme score have the highest density of terms containing chemical formulas (e.g. ``MgB$_2$''). The data points of the two plots are the same as in plot A of Fig.~\ref{score}, but colors showing the relative density of the specific type of terms. Note that the colors are log scaled: white stands for 100\% terms of the given type, yellow for $\approx\,$30\%, orange for $\approx\,$10\%, light red for $\approx\,$2\%, dark red for $<\,$1\%, and black for $<\,$0.1\%.
}
\end{figure*}

The two plots of Fig.~\ref{dens} show the same data points as plot A of Fig.~\ref{score} but with colors standing for the relative density of Wikipedia terms (A) and of terms containing chemical formulas (B). Plot A confirms that phrases in the area of a high meme score (towards the top right) tend to show up as titles of Wikipedia articles on physics. Additionally, the plot shows that this is the \emph{only} such area. There are a few scattered outliers, but the only significant area with a high density of Wikipedia terms is found around the 99.9\%-quantile.
Plot B shows that phrases containing chemical formulas (such as ``BaFe$_2$As$_2$'') tend to have a relatively low frequency (individually) but high propagation score. The area with the highest density can again be found along the 99.9\%-quantile, which is consistent with the expectation of chemical compounds to be important and interesting entities for physics research. The fact that they are standardized and compressed representations reduces moreover their ``vulnerability'' to synonyms or spelling variants, making them stronger memes on the level of pure character sequences. (We come back to the issue of memes on different levels of abstraction below.)

\begin{figure*}[tb]
\centering{\includegraphics[width = \textwidth]{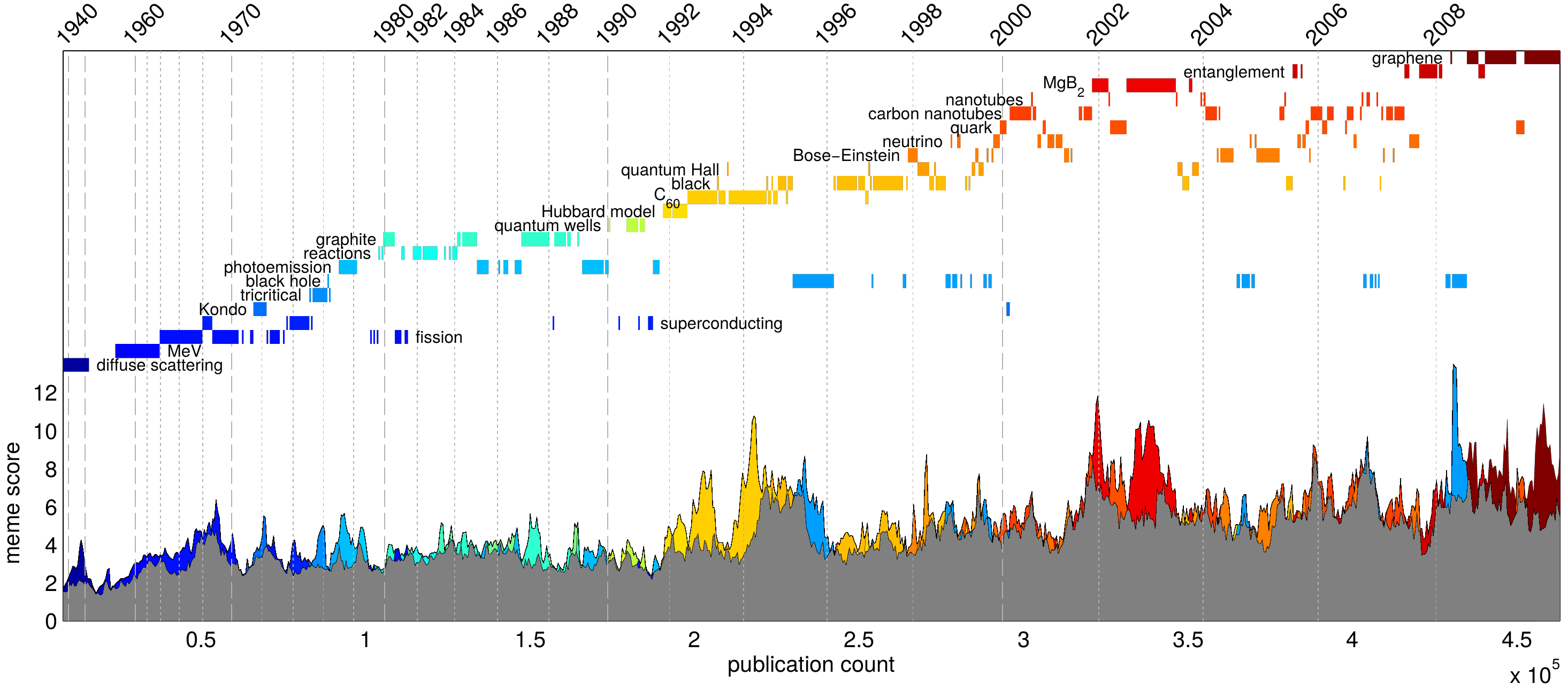}}
\caption{
\label{time}
Time history of top physics memes based on their meme scores obtained from the American Physical Society dataset. The time axis is scaled by publication count. Bars and labels are shown for all memes that top the rankings for at least ten out of the displayed $911$ points in time. The gray area represents the second-ranked meme at a given time.
}
\end{figure*}

Fig.~\ref{time} shows the top memes over time, revealing bursty dynamics, akin to the one reported previously in humans dynamics \cite{barabasi_n05} and the temporal distribution of words \cite{altmann_pone09}. These bursts might be a reflection of the fast rise and fall of many scientific memes in terms of their popularity. As new scientific paradigms emerge, the old ones seem to quickly lose their appeal, and only a few memes manage to top the rankings over extended periods of time. The bursty dynamics also support the idea that both the rise and fall of scientific paradigms is driven by robust principles of self-organization \cite{perc_sr13}.

\section*{Discussion}

By going back to the original analogy with genes put forward by Richard Dawkins \cite{dawkins_89}, we investigated the relation between the occurrence frequency of scientific memes and the degree to which they propagate along the citation graph. We found that scientific memes are indeed governed by a surprisingly simple relationship of these two factors.
This is formalized by the meme score --- a metric to characterize and identify scientific memes --- defined as the product of the frequency of occurrence and the propagation score.

We have shown that the meme score can be calculated exactly and exhaustively without the introduction of arbitrary thresholds or filters and without relying on any kind of linguistic or ontological knowledge. The method is fast and reliable, and it can be applied to massive databases. We have demonstrated the effectiveness of the meme score on more than 47 million publication records from the Web of Science, PubMed Central, and the American Physical Society. Moreover, we have evaluated the accuracy of the proposed meme score by means of full and time-preserving randomizations of the citation graphs, by means of manual annotation of publications, as well as by means of several alternative metrics. We have provided statistical evidence for the agreement between human annotators and the meme-score results, and we have shown that it is superior to alternative metrics. We have also confirmed that the observed patterns cannot be explained by topological or temporal features alone, but are grounded in more intricate processes that determine the dynamics of the scientific progress and the way credit is given to preceding publications.
Furthermore, the top-ranking scientific memes reveal a bursty time dynamics, which might be a reflection of the fierce competition among memes for the limited and fluctuating resource of scientists' attention.

We have only considered fixed character sequences as potential memes, but it is clear that memes do not only exist on this low level, and it is reasonable to expect that the inclusion of additional layers of processing using linguistic and ontological resources would lead to even better results and would let us capture memes on a more abstract level. Such memes might consist of sets of morphological variants, co-occurrences of words, compositions of multiple memes, grammatical constructions, or even argumentation schemes and rhetorical styles.
We deliberately kept the meme score as simple as possible to emphasize that it is surprisingly precise on its own. At the same time, there are many ways to improve the metric in the future with more sophisticated processing to capture memes on a higher level.
In general, we believe that the presented approach by allowing to study memes in a comprehensive manner opens up the field for a wide range of future research on topics such as information diffusion, complex systems, innovation, scientific progress, social dynamics, ecosystems, cultural evolution, and of course the study of memes themselves.

\section*{Methods}

\subsubsection*{Graph randomization}

The analyzed randomized networks have exactly the same topology as the original ones but the article texts (i.e. titles and abstracts with their memes) are randomly assigned to the nodes. Each node therefore owes its position in the network to one particular publication but has text attached that comes from a different one.
For the time-preserving randomizations, we shuffle only publications that were published within narrow consecutive time windows. Concretely we use time windows of 1000 publications, meaning that --- after shuffling --- no publication has moved more than 1000 positions forward or backward from the original chronological order.

\subsubsection*{Human annotation}

For the first part of the manual annotation, we use the following two categories: (i) the phrase is not a meaningful term or not an important concept of physics, and (ii) the phrase is an important concept or entity of physics --- it could appear as the title of an entry of a comprehensive encyclopedia of physics. For the second part, we defined the following linguistic phrase types: (i) noun phrase, (ii) verb, (iii) adjective or adverb, and (iv) other.

The set of phrases used for this evaluation consisted of the top 150 memes with respect to their meme score, extracted from the American Physical Society dataset, plus another two sets for comparison of 150 randomly drawn phrases each. For the two comparison sets, we considered all phrases that appear in at least 100 publications. From these, 150 terms were drawn randomly without taking into account their frequency, i.e., frequent terms had the same chance of being selected as infrequent ones, whereas the 150 terms of the second set were drawn with a weight that corresponded to their frequency.
Moreover, to rule out effects of different $n$-gram lengths, we made sure that the two batches of random terms followed exactly the same length distribution as the main sample extracted based on the meme score. The resulting 450 terms were shuffled and given to two human annotators, both PhD students with a degree in physics, who independently annotated the terms.

\subsubsection*{Metrics for comparison}

We have used the following metrics with different parameter settings as alternatives to the meme score (see Supplementary Material for details):
(i) frequency --- the most frequent terms, optionally skipping the first $x$ terms; (ii) maximum absolute change over time --- the highest-scoring terms with respect to maximum \emph{absolute} change in frequency; (iii) maximum relative change over time --- the same as (ii) but based on \emph{relative} changes; (iv) maximum absolute difference across journals --- the highest-scoring terms with respect to maximum \emph{absolute} difference in frequency between journals; (v) maximum relative difference across journals --- the same as (iv) but based on \emph{relative} changes.
Metric (i) is based on the assumption that important memes are frequent but not as frequent as the small class of general words that can be found in all types of texts. Metrics (ii) and (iii) are based on an idea proposed in \cite{perc_sr13}, being that interesting memes exhibit trends over time.
Metrics (iv) and (v) are based on the intuition that phrases occurring mostly in specific journals but not in others must be specific concepts of the particular field of research.

As a ground-truth list of memes, we automatically extracted $5178$ terms from Wikipedia. We collected the titles of all articles --- and terms redirecting to them --- from the categories ``physics'', ``applied and interdisciplinary physics'', ``theoretical physics'', ``emerging technologies'', and their direct sub-categories, but filtering out terms that appear in less than $10$ publications of the American Physical Society dataset.
To quantify the agreement between the top memes identified by a particular metric and the Wikipedia list, we use the normalized area $A$ under the curve as shown on Fig.~\ref{wiki}. The step-shaped curved has a log-scaled $x$-axis running up to the number of terms $s$ on the ground-truth meme list ($s = 5178$ in our case) and an $y$-axis running from $0$ (no overlap) to $1$ (perfect overlap).
Limiting cases are $A = 1$, representing perfect agreement, and $A = 0$, representing no agreement at all between the two compared lists.

\begin{acknowledgments}
This research was supported by the European Commission through the ERC Advanced Investigator Grant ``Momentum'' (Grant No. 324247) and by the Slovenian Research Agency through the Program P5-0027.
In addition, we would like to thank Karsten Donnay, Matthias Leiss, Christian Schulz, and Olivia Woolley-Meza for their useful feedback and help with the realization of the evaluations.
\end{acknowledgments}

\bibliography{memes}
\end{document}